\documentclass[a4paper]{jpconf}
\usepackage{graphicx}
\newcommand {\be}{\begin{equation}}
\newcommand {\ee}{\end{equation}}
\newcommand {\bea}{\begin{eqnarray}}
\newcommand {\eea}{\end{eqnarray}}

\begin{document}
\title{Weakening of antiferromagnetism due to frustration on inhomogeneous 2D lattices}

\author{Anuradha Jagannathan}

\address{Laboratoire de Physique des Solides, Universit\'e Paris-Sud, 91405 Orsay, France}

\ead{jagannathan@lps.u-psud.fr}

\begin{abstract}
I consider the ground state of quantum spins interacting via
Heisenberg antiferromagnetic exchange, in lattices having several
types of local environment. In unfrustrated cases, when a N\'eel
type order exists, it is known that quantum fluctuations are
site-dependent, and lead to an inhomogeneous local staggered order
parameter. This paper discusses some effects of adding frustration
to such a system, by including next nearest neighbor
antiferromagnetic interactions. Within linear spinwave theory, it is
found that as the frustration parameter increases, the local order
parameters decrease at a rate that depends on the type of sites
involved in the frustrated loops. The cases of a decorated square
lattice system, and of a quasiperiodic tiling are considered to
study different ways in which antiferromagnetism is suppressed in
these inhomogeneous systems.
\end{abstract}

\section{Introduction}
The work reported here is a study of the different means of
suppressing antiferromagnetic order by frustration, in systems with
several local environments for spins. The calculations are carried
out within spin wave theory. Their starting point is the N\'eel
ordered ground state of $S=\frac{1}{2}$ Heisenberg models in
unfrustrated two dimensional (2D) lattices in which all couplings
are short-range, antiferromagnetic and of equal magnitude, denoted
by $J$. The structures I consider are inhomogeneous in that their
unit cell contains sites having different coordination numbers. The
classical antiferromagnetic ground state for such systems has a
collinear two-sublattice N\'eel ordering with a staggered local
magnetization $m_{si} = \epsilon_i \langle S_i^z\rangle =
\frac{1}{2}$ uniformly on all sites, where $\epsilon_i =\pm 1$
depending on whether the site $i$ belongs to the A(B) sublattice,
and $z$ is the axis of symmetry breaking. The quantum ground state
is inhomogeneous \cite{jwm}, with varying $m_{si}$ values. The
staggered moment tends to be large, for example, on sites with
smallest coordination number $z$ (not to be confused with the
spatial z-direction) \cite{note}.

When additional frustrating interactions are added to the original
Heisenberg Hamiltonian via J' exchange terms that
antiferromagnetically couple spins belonging to the same sublattice,
both the global staggered moment, as well as the local order
parameters will eventually be suppressed for large enough $J'$, when
the system goes over to a new state: a coplanar state, or
nonmagnetic spin liquid state, for example. Spin wave theory, which
works well for small $J'$ will also break down before the transition
occurs. Results will be presented for a range of values $0<J'<J'_u$,
where $J'_u$ is the value of the exchange interaction $J'$ for which
linear spin wave theory becomes unstable.

I discuss first a simple square lattice ferrimagnetic example, in
which the sublattices are of unequal size (i.e. the number of sites
in the A-sublattice, $N_A$ is half that of the B-sublattice $N_B = 2
N_A$) to illustrate this effect. I will then consider a more
complicated example of a quasiperiodic case with two equal
sublattices, which has an $S=0$ collinear N\'eel type ground state.

\begin{figure}[ht]
\begin{center}
\includegraphics[scale=0.50]{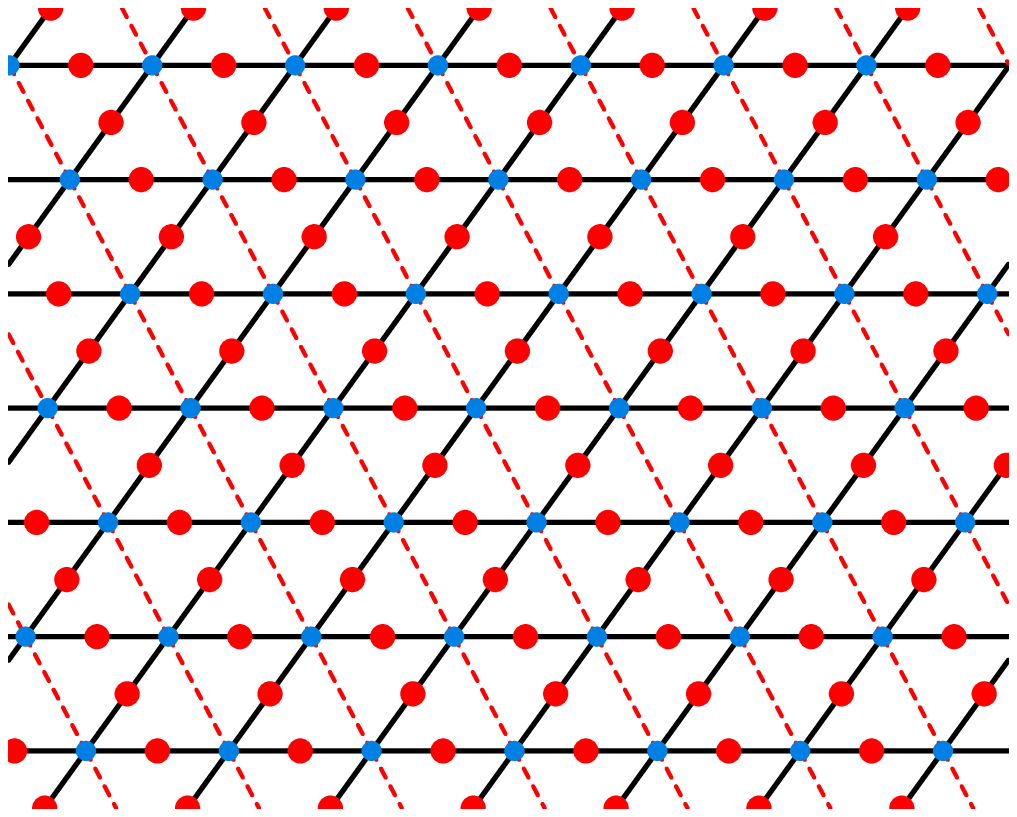}
\hskip 1cm
\includegraphics[scale=0.50]{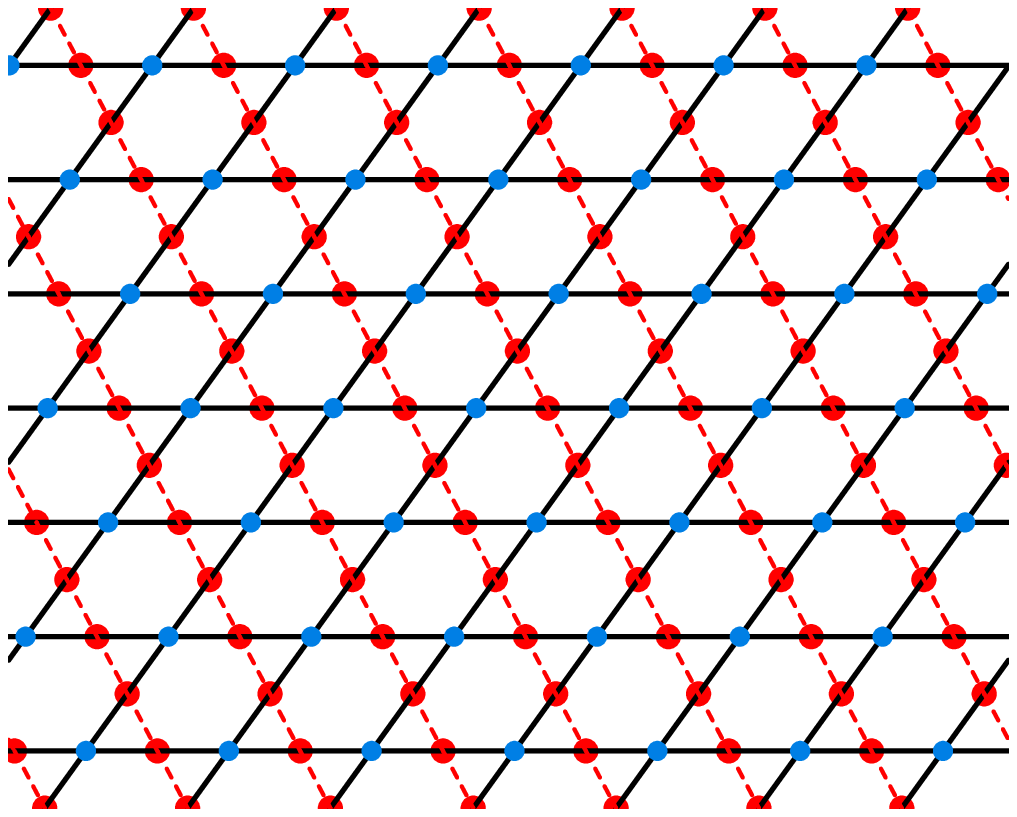}
\vspace{.2cm} \caption {Decorated square lattice system showing the
$J$ bonds (blue) and $J'$ bonds (dashed red) for cases I (left) and
II(right). Sites of the A(B) sublattice spins are shown in blue
(red).} \label{latticefigs.fig}
\end{center}
\end{figure}

\section{The decorated square lattice}
The unperturbed Heisenberg Hamiltonian is

\be H = J\sum_{\langle i,j\rangle} \vec{S}_i.\vec{S}_j \ee

where the spins are placed on the vertices of the square lattice
(the A-sublattice), and on midpoints of each of the sides of the
squares (the B-sublattice). The exchange interaction $J$,
represented in blue in Figs.1, gives rise to a staggered arrangement
of spin up on the A-sublattice and spin down on the B-sublattice.
Since the number of sites in the two sublattices is unequal, there
will be a net magnetization, and this is reflected in the magnon
spectrum. Low energy modes have energy increasing as the square of
the wave number, $\omega(k) \sim k^2$ instead of linearly in $k$, as
for the example that will be discussed next.

In linear spin wave theory, the quantum corrections for the onsite
staggered magnetization (see \cite{jwm} for a description of similar
models) gives $ m_{s}^{(A)} \approx 0.28$ and $ m_{s}^{(B)} \approx
0.38 $ for the two sublattices. The large difference of values of
sublattice magnetization in this (and similar models) is in keeping
with results of Quantum Monte Carlo simulations. The difference can
be explained by an argument based on the Heisenberg star cluster
\cite{jwm} -- consisting of a central spin connected to $z$ nearest
neighbors -- for which the quantum fluctuations about a N\'eel
ordered state are $z$-fold stronger for the center spin compared to
the spins on the extremities. This qualitatively explains the
smaller staggered moment on the A-sublattice sites (coordination
number $z = 4$) compared to the B-sublattice sites (for which
$z=2$).

We now consider two different ways of introducing frustration, which
are illustrated in Figs.1a and b). In case I, bonds are added along
the diagonals of the square lattice: the A-sublattice spins are
coupled via a new term $J' \vec{S}_i.\vec{S}_j$. For small enough
$J'$, such that the symmetry of the ground state is not changed,
this leads to a change in the classical ground state energy per site
of $\Delta E_0/N = J'S^2/3$. Note: in the figures, bond angles and
distances have been distorted, to emphasize the relation to the
triangular lattice system, but not the topological connectivity,
which remains that of the square lattice system when $J'=0$.

In case II, the new terms link spins lying on the B-sublattice,
($i,j \in B$). The change in ground state energy per site in this
case is given by $\Delta E_0/N = 2J'S^2/3$. The parameter $f =
\Delta E_0/NS^2$ will be used below as a measure of the frustration
in each of the two cases considered.

Although our calculations are valid for a large range of $J'$, we
note that our concept of slightly perturbing the original topology
is only valid for small $J'$. As $J'$ becomes large, case II, for
example, is better considered as an anisotropic Kagome system, since
all sites have the same connectivity.

\subsection{Case I. Frustration via coupling between high $z$ lattice sites}
As $J'$ increases, the staggered magnetization decreases on both
sublattices, as one can see from the red (sublattice A) and blue
(sublattice B) curves in Fig.2 which are marked Case I. The f
parameter is given by $f=J'/3$ for these two curves, which extend
out to a value of $f$ determined by the point at which linear spin
wave theory around the collinear state becomes unstable.

\subsection{Case II. Frustration via coupling between low $z$ lattice sites}
As $J'$ increases, the staggered magnetization decreases on both
sublattices, as one can see from the red (sublattice A) and blue
(sublattice B) curves in Fig.2 which are marked Case II. The f
parameter is defined by $f=2J'/3$ for these two curves. Linear spin
wave theory becomes unstable in this case for a value of $f\approx
0.33$.  As can be seen in the figure, the changes in the staggered
order parameters as a function of f are relatively smaller in this
case.

\begin{figure}[ht]
\begin{center}
\includegraphics[scale=0.50]{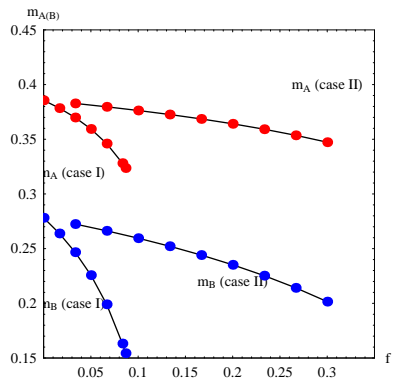}
\vspace{.2cm} \caption {Plot of staggered magnetization for the A
and B sublattice sites for cases I and II } \label{results.fig}
\end{center}
\end{figure}

This example shows the influence of the local environment:
frustration effects can be greater or smaller depending on the type
of sites linked by the $J'$ terms. In this example, as $J'$ is
``turned on", one finds a larger suppression of the global and local
order parameters for the situation where the high-$z$ sites are
involved. For these sites, the big quantum fluctuations induced by
large nearest neighbor number are enhanced by the frustrating terms.
The next example considers frustration in a more complex model in
which A- and B-sublattices are equivalent, and each one has many
different local environments.

\section{Penrose tiling antiferromagnet}

\begin{figure}[ht]
\begin{center}
\includegraphics[scale=0.40]{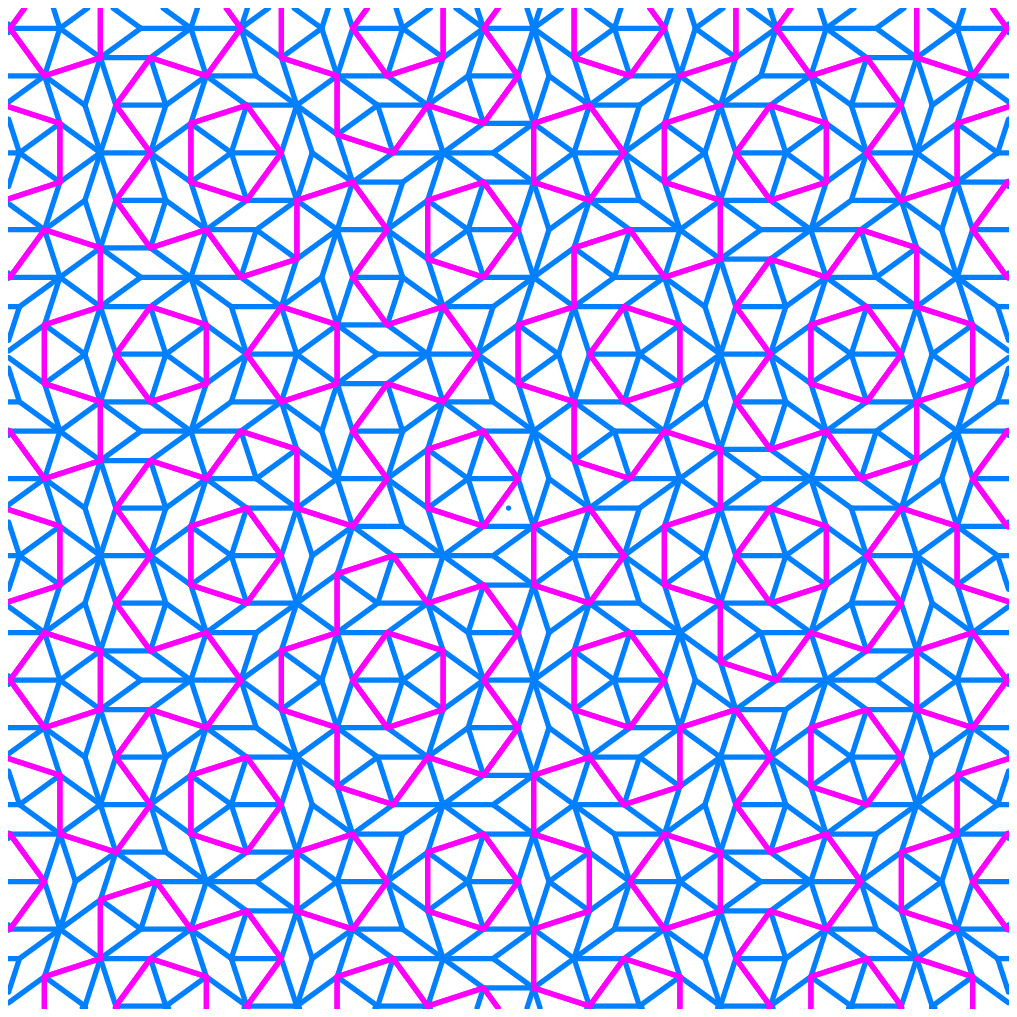}
\includegraphics[scale=0.40]{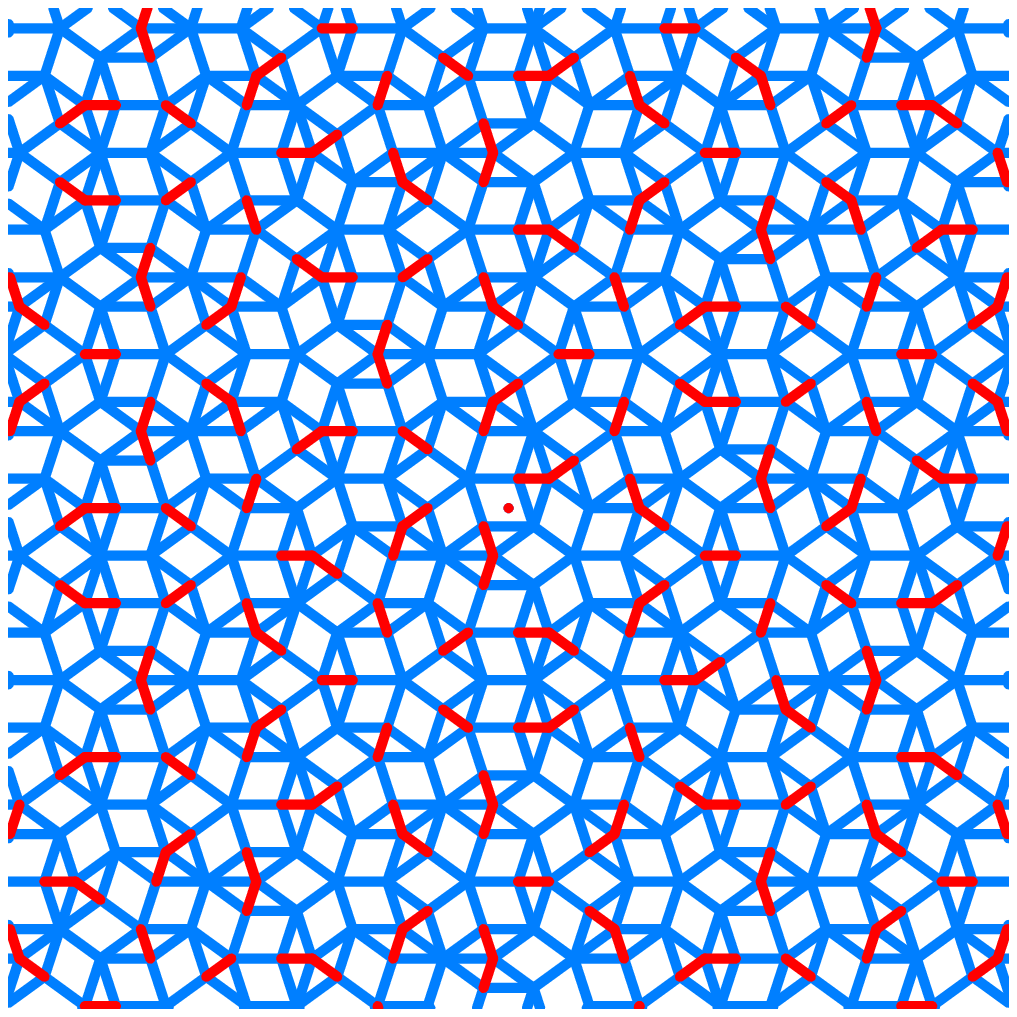}
\includegraphics[scale=0.70]{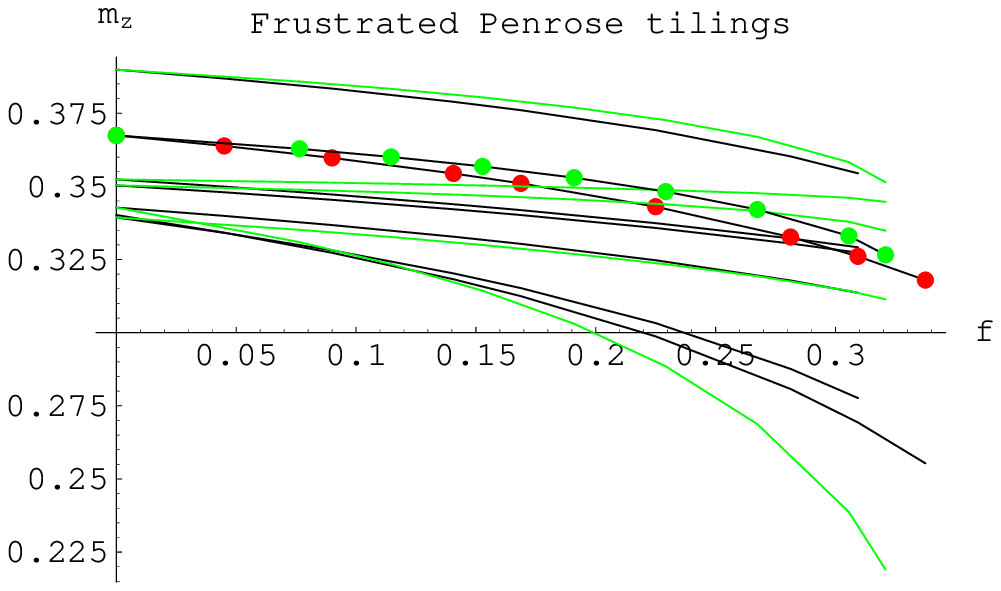}
\vspace{.2cm} \caption { Frustrated Penrose tilings for case III
(left) and case IV (middle) showing the $J$-bonds in blue and the
$J'$ bonds in red. (right)Plots of the staggered magnetization
averaged over all sites (red points: case III, green: case IV) and
of different subsets of sites as a function of $f$ }
\label{resultpenrose.fig}
\end{center}
\end{figure}

The Penrose rhombus tiling, considered here, is bipartite with two
equivalent sublattices. Numerical calculations of the spin wave
spectrum were carried out for finite systems with $N_A=N_B$, which
possess an $S=0$ ground state, for the unperturbed $J'=0$
Hamiltonian. The figures \ref{resultpenrose.fig} show the strong $J$
bonds in blue. These bonds act along the tile edges of the two
building blocks of the Penrose tiling, the $36^\circ$ -- or thin --
rhombus and the $72^\circ$ -- or thick -- rhombus).  The new
frustration-inducing diagonal $J'$ bonds are shown in red for two
different situations. Case III corresponds to connecting sites
across the diagonal of the thick rhombuses, while case IV
corresponds to connecting sites on the diagonals of the thin
rhombuses. Note that these bonds connect different classes of sites,
and affect A- and B-sublattices equally. The $J'$ bonds frustrate
sites to different degrees, the classification of sites into
families depending largely on their coordination number, which
ranges from 3 to 7. An inspection of the figures will make it clear
that, for example, the connectivities of $z=6$ and $z=7$ vertices
are not modified. In brief, the $J'$ terms have their strongest
effects on a subset of $z=5$ sites for the case III and $z=4$ sites
in case IV.

The frustration parameter $f$, determined by the shift in the
classical ground state energy as in the previous square lattice
example, here takes the values $f=0.56 J'$ (case III) and $f=0.38
J'$ (case IV) respectively.

The spin wave spectrum and eigenmodes are calculated by the same
technique that was used for the unfrustrated problem (see
\cite{prbrapid}, namely, via a real space numerical diagonalization
of the linearized spin wave Hamiltonian. The theory breaks down for
large enough $J'$, as in the preceding examples. Although a detailed
analysis is still in progress, some results for the local staggered
magnetizations are indicated in Fig. \ref{resultpenrose.fig}. The
figure shows data for a system of 1686 spins in each of the two
cases. For cases III and IV, the globally averaged staggered
magnetization is shown, in red (III) and green (IV), as a function
of $f$. The remaining curves show the decay of the average
sublattice magnetization within each family, for the two problems.
In each case, it is observed that one particular subset of sites
reacts the most strongly to the perturbation. These subsets are the
previously-mentioned $z=5$ (case III) and $z=4$ sites (caseIV), and
they are also the sites of maximal quantum fluctuations in the
absence of frustration. In these two examples of frustrating the
Penrose tiling, the frustration acts to weaken the N\'eel order in
different ways locally, with the same global result for the
staggered magnetization. It will be interesting to explore different
ways to add frustration, that involve, for example the higher
coordination number sites ($z=6,7$), to see the effect on the decay
of collinear magnetic order.

\section{Conclusions} In inhomogeneous systems, quantum fluctuations suppress the local staggered
magnetizations on certains subsets of sites. Several examples are
presented here of the role of frustration in suppressing the N\'eel
order in such systems. While the details of the suppression
mechanism probably depend on the specific lattice structure, the
trend observed here indicates more efficient suppression when
frustration is added on highly connected sites.





\smallskip


\medskip
\begin{thebibliography}{9}
\bibitem{jwm} Jagannathan A, Wessel S and Moessner R 2006 {\it Phys. Rev.} B {\bf 74} 184410
\bibitem{note} The dependence of
$m_{si}$ on coordination number is not necessarily monotonic, since
next nearest neighbor effects are also relevant. \bibitem{prbrapid}
Szallas A, Jagannathan A, Wessel S and Duneau M 2007 {\it Phys.
Rev.} B {\bf 75} 212407; Szallas A and Jagannathan A 2008 {\it Phys.
Rev.} B {\bf 77} 104427
\end{thebibliography}
\end{document}